\documentclass[twocolumn,showpacs,preprintnumbers,amsmath,amssymb,times,pre]{revtex4}
\usepackage{txfonts}
\usepackage{amssymb}

\usepackage{graphicx}
\usepackage{dcolumn}
\usepackage{bm}
\usepackage{amsmath}

\begin{document}
\title{Optimal View Angle in Collective Dynamics of Self-propelled Agents}

\author{Bao-Mei Tian$^{1}$}
\author{Han-Xin Yang$^{1}$}
\author{Wei Li$^{2}$}
\author{Wen-Xu Wang$^{4}$}
\author{Bing-Hong Wang$^{1,5}$}
\author{Tao Zhou$^{1,3}$}

 \affiliation{$^{1}$Department of Modern Physics, University of
 Science and Technology of China, Hefei 230026, China\\
 $^{2}$Department of Automation, Shanghai Jiao Tong University, Shanghai 200240, China\\$^3$Department of Physics, University of Fribourg, Chemin
 du Mus\'ee 3, CH-1700 Fribourg, Switzerland\\$^{4}$Department of
Electronic Engineering, Arizona State University, Tempe, Arizona
 85287-5706, USA\\$^{5}$The Research Center for Complex System Science, University of Shanghai for Science
and Technology, Shanghai 200093, China}

\date{\today}

\begin{abstract}
We study a system of self-propelled agents with the restricted
vision. The field of vision of each agent is only a sector of disc
bounded by two radii and the included arc. The inclination of
these two radii is characterized by the view angle. The
consideration of restricted vision is closer to the reality
because natural swarms usually do not have a panoramic view.
Interestingly, we find that there exists an optimal view angle,
leading to the fastest direction consensus. The value of the
optimal view angle depends on the density, the interaction radius,
the absolute velocity of swarms and the strength of noise. Our
findings may invoke further efforts and attentions to explore the
underlying mechanism of the collective motion.
\end{abstract}

\pacs{05.60.Cd, 87.10.-e, 89.75.Hc, 02.50.Le}

\maketitle

The collective motion of a group of autonomous agents (or particles)
\cite{ref1,ref2,ref3,ref4,ref5,ref6,ref7,ref8} has attracted much
attention in the past decade. One of the most remarkable
characteristics of systems, such as flocks of birds, schools of
fish, and swarms of locusts, is the emergence of collective states
in which the agents move in the same direction. A particularly
simple and popular model to describe such behavior was proposed by
Vicsek \emph{et al.} \cite{Vicsek}. Due to simplicity and
efficiency, the Vicsek model (VM) has been intensively investigated
in recent years
\cite{recent1,recent2,recent3,recent4,recent5,recent6,recent7,recent8,recent9,recent10,recent11,recent12,recent13}.

In the VM, $N$ agents move synchronously in a square shaped cell of
linear size $L$ with the periodic boundary conditions. The initial
directions and positions of the agents are randomly distributed in
the cell, and each agent has the same absolute velocity $v_{0}$.
Agent $i$ and agent $j$ are neighbors at time step $k$ if and only
if $\parallel \vec{X}_{i}(k)-\vec{X}_{j}(k)\parallel \leq R$, where
$\vec{X}_{i}(k)$ denotes the position of agent $i$ on a
2-dimensional (2D) plane at time step $k$ and $R$ is the sensor
radius. The direction of agent $i$ at time step $k+1$ is:
\begin{equation}
\theta_{i}(k+1)=\langle \theta_{i}(k) \rangle_{R}+\Delta \theta,
\end{equation}
where $\langle \theta_{i}(k)\rangle_{R}$ denotes the average
direction of agent $i$'s neighbors (include itself), $\Delta \theta$
denotes noise (in the following discussions, $\Delta \theta=0$
without special mention). To be more specific, let $\Gamma_{i}(k)$
be the set of neighbors of agent $i$ at time step $k$, the VM is
then described as \cite{recent7,recent8}:

\begin{equation}
\vec{X}_{i}(k+1) = \vec{X}_{i}(k)+ v_{0}e^{i\theta_{i}(k)}\Delta t,
\end{equation}
\begin{equation}
\theta_{i}(k+1)=angle(\sum_{j\in\Gamma_{i}(k+1)}e^{i\theta_{j}(k)}),
\end{equation}
where $e^{i\theta_{i}(k)}$ is the unitary complex directional vector
of agent $i$,
$e^{i\theta_{i}(k)}=\cos(\theta_{i}(k))+i\sin(\theta_{i}(k))$,
$\theta_{i}(k)\in[0,2\pi)$. Here the function $angle(\cdot)$ denotes
the angle of a complex number. $\theta_{i}(k+1)$ is the moving
direction of agent at time step $k+1$, which is the average
direction of agents in the neighbor set $\Gamma_{i}(k+1)$.
$v_{0}e^{i\theta_{i}(k)}$ represents the velocity of agent $i$ at
time step $k$ with constant speed $v_{0}$ and direction
$\theta_{i}(k)$.

In the VM and most other models of self-propelled particles, the
field of vision for every agent is a complete disc (2D case) or a
sphere (3D case) characterized only by its sensor radius $R$. In the
reality, however, most animals are incapable of complete view. For
example, the cyclopean retinal field of human is about 180 degree
and the cyclopean retinal field of tawny owl is 201 degree
\cite{Martin}. It is thus more reasonable to assume limited view
angles of agents \cite{ref3,Huth}, instead of the omnidirectional
views, in swarm models to better mimic the real collective
behaviors.

\begin{figure}
\includegraphics[width=0.53\textwidth]{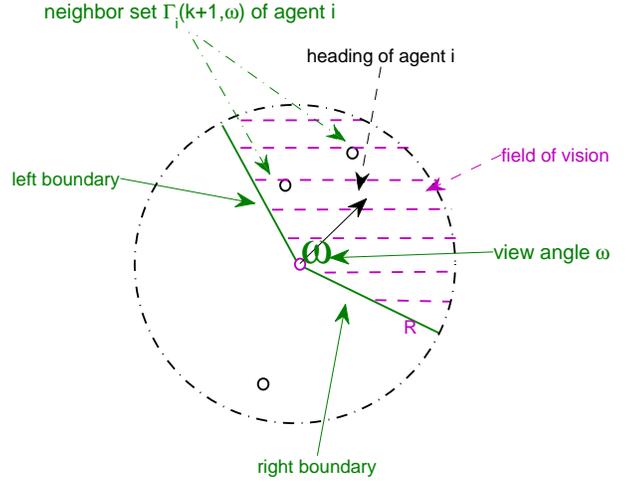}
\caption{\label{fig:epsart}(Color online) Illustration of the
non-omnidirectional view of agent $i$ at time step $k+1$ in a 2D
plane.}
\end{figure}

\begin{figure}
\scalebox{0.8}[0.8]{\includegraphics{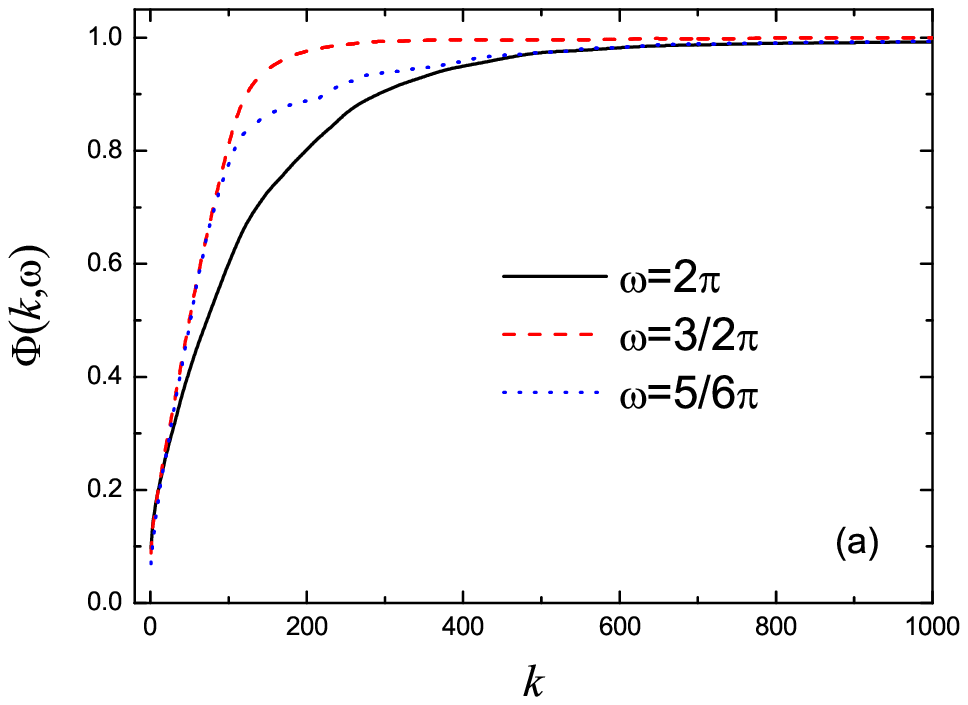}}
\scalebox{0.8}[0.8]{\includegraphics{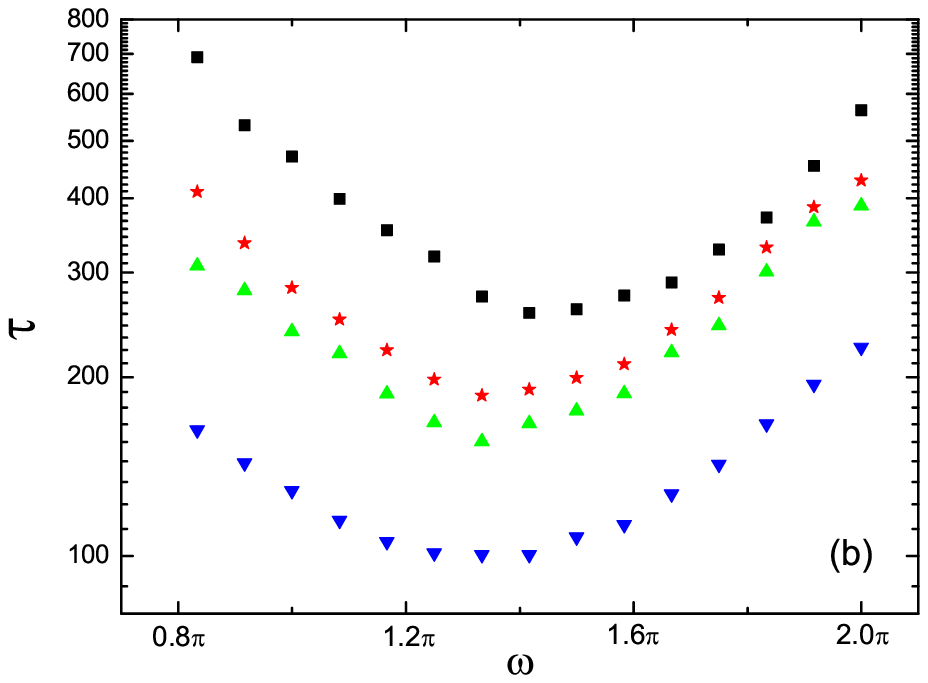}}
\caption{\label{fig:epsart} (Color online) (a) The order parameter
$\Phi(k,\omega)$ as a function of time step $k$ for different values
of view angle $\omega$. Here $N$=400, $R$=0.6, $v_{0}=0.04$. (b) The
transient time step $\tau$ as a function of the view angle $\omega$.
The symbols correspond to $\blacksquare$: $R$=0.6, $v_{0}=0.02$,
$N$=400; $\bigstar$: $R$=0.6, $v_{0}=0.04$, $N$=400;
$\blacktriangle$: $R$=0.6, $v_{0}=0.04$, $N$=500;
$\blacktriangledown$: $R$=0.8, $v_{0}=0.04$, $N$=400. Each data
point is obtained by averaging over 500 different realizations.}
\end{figure}

In this Brief Report, we investigate the VM in which agents have
limited view angles $\omega$, $\omega\in(0,2\pi]$. As illustrated in
Fig. 1, the field of vision of every agent is only a sector of disc
bounded by two radii and the included arc, the left (right) boundary
of vision and the heading of agent $i$ have inclination $\omega/2$,
that is, for every agent, the field of view is symmetric about its
current moving direction. Thus rule (3) in the VM can be modified
as:
\begin{equation}
\theta_{i}(k+1)=angle(\sum_{j\in\Gamma_{i}(k+1,\omega)}e^{i\theta_{j}(k)}),
\end{equation}
where $\Gamma_{i}(k+1,\omega)$ denotes the neighbor set of agent $i$
with view angle $\omega$. When $\omega = 2\pi$, the rule (4)
degenerates to the original Vicsek model (3).

To give a quantitative discussion, we define an order parameter
\begin{equation}
\Phi(k,\omega)=\frac{1}{N}|\sum_{i=1}^{N}e^{i\theta_{i}(k)}|,  0\leq
\Phi(k,\omega)\leq1
\end{equation}
for the system (4) at time step $k$ with view angle $\omega$,
obviously, $0 \leq \Phi(k,\omega) \leq 1$.

In noiseless case, the order parameter $\Phi(k,\omega)$ can approach
1 when the evolution is long enough, except for extremely rare cases
(for example, the cases may occur when $R$ or $\omega$ is too
small). To quantify the speed of direction consensus, we study the
transient time step $\tau$, which is defined as the time step when
the order parameter first surpasses a certain value $\Phi_{0}$. Here
we take $\Phi_{0} = 0.99$ and we have checked that qualitative
results are not changed when $\Phi_{0}$ is large enough.

We then investigate the effects of the view angle $\omega$ on the
transient process. As shown in Fig. 2(a), the order parameter
$\Phi(k,\omega)$ reaches 1 faster when the view angle
$\omega=3\pi/2$, compared with $\omega=2\pi$ and $\omega=5\pi/6$.
Figure 2(b) shows the transient time step $\tau$ as a function of
$\omega$ for different values of parameters. One can find that
$\tau$ is not a monotonic function of $\omega$ and there exists an
optimal view angle, leading to the shortest transient time.

\begin{figure}
\scalebox{0.75}[0.8]{\includegraphics{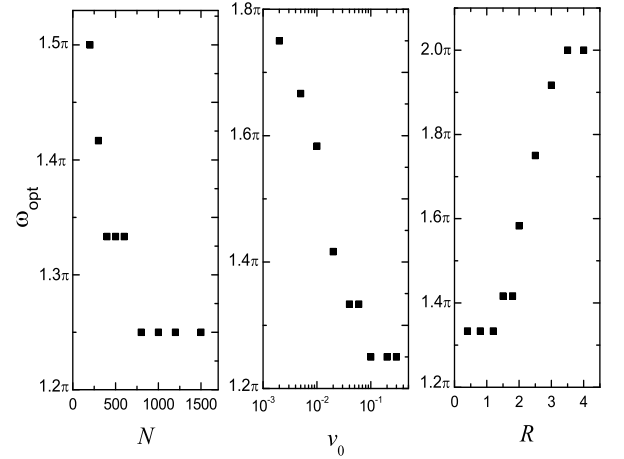}}
\caption{\label{fig:epsart} The optimal view angle $\omega_{opt}$ as
functions of the swarm number $N$, sensor radius $R$ and absolute
velocity $v_0$ respectively. For left panel: $R$=0.6, $v_{0}$=0.04;
for middle panel: $R$=0.6, $N$=400 and for right panel:
$v_{0}$=0.04, $N$=400. The lattice size is fixed as $L=10$. Each
data point is obtained by averaging over 500 different realizations.
Note that, the resolution of view angle in our simulation is set to
be $\pi/12$. }
\end{figure}

Figure 3 shows the optimal view angle $\omega_{opt}$ as functions
of the parameters: the swarm number $N$, the sensor radius $R$ and
the absolute velocity $v_0$, respectively. One can see that the
optimal view angle $\omega_{opt}$ decreases with the increasing of
$N$ and $v_0$, and converges to a fixed value when $N$ or $v_0$ is
large enough. $\omega_{opt}$ increases as the sensor radius $R$
increases. In particular, when $R$ being close to the lattice size
$L$, agents with panoramic view will be globally coupled and the
directions of the swarm can reach consensus in only one time step.

We next investigate whether more communications are needed for
faster convergence. We define $n_{i}(k,\omega)$ as the number of
$i$'s neighbors, and the average number of neighbors $\langle
n(k,\omega) \rangle$ over all agents at time step $k$ is
\begin{equation}
    \langle n(k,\omega) \rangle = \frac{1}{N} \sum\limits_{i = 1}^N {n_i (k,\omega
    )}.
\end{equation}

In Fig. 4, we report this average neighboring number for different
$\omega$. Combining Fig. 2(a) and Fig. 4, it is interesting to find
that, agents with optimal view angle $\omega=3\pi/2$ have the least
number of neighbors in the steady state, compared with
$\omega=2\pi$, $\omega=5\pi/6$ and $\omega=\pi$. This result
indicates the existence of superfluous communications in the VM,
which may counteract the direction consensus.

\begin{figure}
\scalebox{0.8}[0.8]{\includegraphics{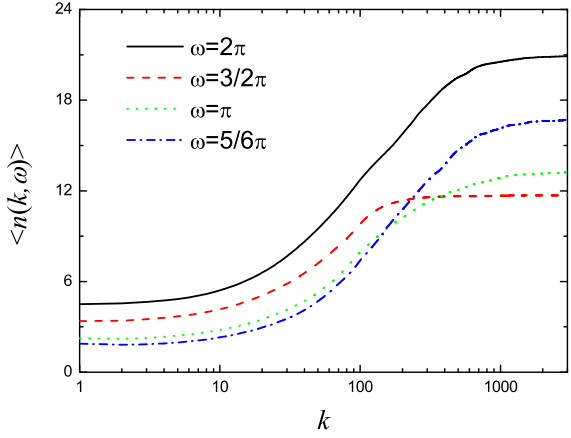}}
\caption{\label{fig:epsart} (Color online) The average number of
neighbors $\langle n(k,\omega) \rangle $ as a function of time step
$k$ for different view angle $\omega$. Here the parameters $N$, $L$,
$R$ and $v_{0}$ are the same with the parameters in Fig. 2(a). Each
data point is obtained by averaging over 500 different realizations.
}
\end{figure}

In the following, we focus on the noise effects associated with the
restriction of view angle. The noise is introduced to the view angle
model as:
\begin{equation}
\theta _i (k + 1) = angle\left( {e^{i\xi } \sum\limits_{j \in \Gamma
_i (k + 1,\omega )} {e^{i\theta _j (k)} } } \right),
\end{equation}
where the moving direction of each agent is perturbed by a random
number $\xi$ chosen with a uniform probability from the interval
$[-\eta,\eta]$. In the presence of noise, the order parameter
$\Phi(k, \omega, \eta)$ will fluctuate and never keep fixed at a
certain value, therefor we adopt a statistically stable order
parameter in terms of $\Phi_{stable}(\omega, \eta)$, which is an
average of the consecutive series of $\Phi(k, \omega, \eta)$ over
many time steps after a sufficiently long transient time. Figure 5
shows that $\Phi_{stable}(\omega, \eta)$ increases as $\omega$
increases if the noise is kept constant, and decreases as the
noise increases.

\begin{figure}
\scalebox{0.8}[0.8]{\includegraphics{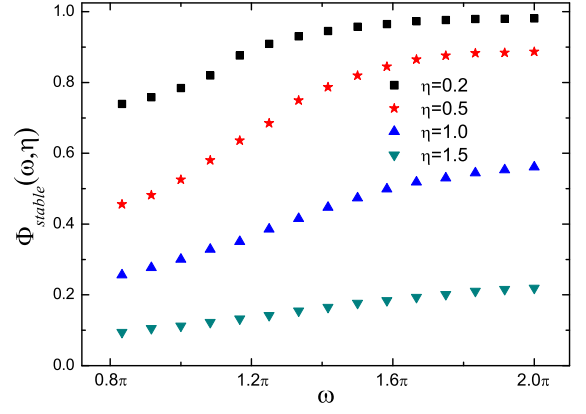}}
\caption{\label{fig:epsart} (Color online) The statistically-stable
order parameter $\Phi_{stable}(\omega, \eta)$ as a function of the
view angle $\omega$ for different noise $\eta$. Here
$\Phi_{stable}(\omega, \eta) = \frac{1}{500}\sum\nolimits_{k =
2501}^{3000} {\Phi (k,\omega,\eta )} $. $N$=400, $L$=10, $R$=0.6,
$v_{0}=0.04$. Each data point is obtained by averaging over 500
different realizations. }
\end{figure}

\begin{figure}
\scalebox{0.8}[0.8]{\includegraphics{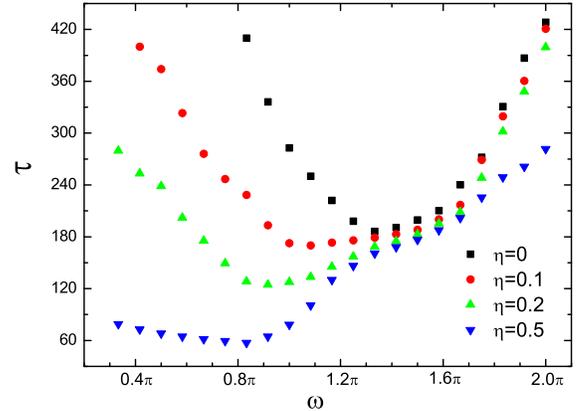}}
\caption{\label{fig:6} (Color online) The transient time step $\tau$
as a function of the view angle $\omega$ for different values of the
noise $\eta$. $N$=400, $L$=10, $R$=0.6, $v_{0}=0.04$. Each data
point is obtained by averaging over 500 different realizations.}
\end{figure}

In the noisy case, we define the transient time step $\tau$ as the
time step when the order parameter firstly exceeds
$0.99\Phi_{stable}(\omega, \eta)$ for each run. For $\eta =0$,
$\Phi_{stable}(\omega, 0)$ approaches 1, thus this definition of
$\tau$ is applicable in the absence of noise. From Fig. 6, one can
find that there still exists an optimal view angle $\omega_{opt}$
leading to the shortest transient time step in the presence of noise
and the value of the optimal view angle decreases as the noise
increases.

In conclusion, we have studied the effects of restricted vision of a
group of self-propelled agents. The field of vision of every agent
is only a sector of disc and the included arc represents the view
angle. It is interesting to find that there exists an optimal angle
resulting in the fastest direction consensus. The value of the
optimal view angle increases as the increasing of sensor radius,
while decreases as the increasing of swarm number, the absolute
velocity or the noise strength. Another interesting phenomenon is
that agents with optimal view angle have the least number of
neighbors in the steady state. Our studies indicate the existence of
superfluous communications in the Vicsek model, which indeed hinder
the direction consensus. Moreover, our results may be useful in
designing the man-made swarms such as autonomous mobile robots.

We thank Hai-Tao Zhang and Ming Zhao for their valuable comments.
This work is funded by the National Basic Research Program of China
(973 Program No.2006CB705500), the National Natural Science
Foundation of China under Grant Nos. 10635040 and 10805045, the
Specialized Research Fund for the Doctoral Program of Higher
Education of China (20060358065).


\begin{thebibliography}{100}

\bibitem{ref1} J. K. Parrish, Science \textbf{284}, 99 (1999).
\bibitem{ref2} H. Levine, W. J. Rappel, and I.
Cohen, Phys. Rev. E \textbf{63}, 017101 (2000).
\bibitem{ref3} I. D. Couzin, J. Krause,
R. James, G. D. Ruxton, and N. R. Franks, J. Theor. Biol.
\textbf{218}, 1 (2002).
\bibitem{ref4} I. D. Couzin, J. Krause, N. R. Franks, and S. A. Levin,
Nature (London) \textbf{433}, 513 (2005).
\bibitem{ref5}  J. Buhl, D. J. T. Sumpter, I. D. Couzin, J. J. Hale, E.
Despland, E. R. Miller, and S. J. Simpson, Science \textbf{312},
1402 (2006).
\bibitem{ref6} M. R. D'Orsogna, Y. L. Chuang, A. L. Bertozzi, and L. S.
Chayes, Phys. Rev. Lett. \textbf{96}, 104302 (2006).
\bibitem{ref7} D. Grunbaum, Science \textbf{312}, 1320 (2006).
\bibitem{ref8} A. Kolpas, J. Moehlis, and I. G. Kevrekidis, Proc. Natl.
Acad. Sci. U.S.A. \textbf{104}, 5931 (2007).


\bibitem{Vicsek} T. Vicsek, A. Czir\'ok, E. Ben-Jacob, I. Cohen, and O.
Shochet, Phys. Rev. Lett. \textbf{75}, 1226 (1995).

\bibitem{recent1} L. Moreau, IEEE Trans. Autom. Control \textbf{50}, 169 (2005).

\bibitem{recent2} F. Cucker and S. Smale, IEEE Trans. Autom. Control \textbf{52},
852 (2007).

\bibitem{recent3} G. Gr\'egoire and H. Chat\'e, Phys. Rev. Lett. \textbf{92}, 025702
(2004).
\bibitem{recent4} C. Huepe and M. Aldana, Phys. Rev. Lett. \textbf{92}, 168701
(2004).
\bibitem{recent5} M. Aldana, V. Dossetti, C. Huepe, V. M. Kenkre, and H.
Larralde, Phys. Rev. Lett. \textbf{98}, 095702 (2007).
\bibitem{recent6} M. Nagy, I. Daruka, and T. Vicsek, Physica A \textbf{373}, 445
(2007).
\bibitem{recent7} W. Li and X. F. Wang, Phys. Rev. E \textbf{75}, 021917
(2007).
\bibitem{recent8} W. Li, H. T. Zhang, M. Z. Q. Chen, and T.
Zhou, Phys. Rev. E \textbf{77}, 021920 (2008).
\bibitem{recent9} W. Li, IEEE Trans. Syst., Man, Cybern. B \textbf{38}, 1084 (2008).
\bibitem{recent10} H. Chat\'{e}, F. Ginelli,
G. Gr\'{e}goire, and F. Raynaud, Phys. Rev. E \textbf{77}, 046113
(2008).
\bibitem{recent11} H. T. Zhang, M. Chen, and T. Zhou, Phys. Rev. E \textbf{79}, 016113
(2009).
\bibitem{recent12} L. Q. Peng, Y. Zhao, B. M. Tian, J. Zhang, B. H. Wang, H. T. Zhang, and T.
Zhou, Phys. Rev. E \textbf{79}, 026113 (2009).

\bibitem{recent13} J. Zhang, Y. Zhao, B. M. Tian, L. Q. Peng, H. T.
Zhang, B. H. Wang, and T. Zhou, Physica A \textbf{388}, 1237 (2009).

\bibitem{Martin} G. R. Martin, J Comp Physiol A \textbf{174}, 787 (1994)

\bibitem{Huth} A. Hutha and C. Wissela, J. Theor. Biol. \textbf{156}, 3 (1992).




\end{thebibliography}
\end{document}